\documentclass[prd,twocolumn,showpacs,reprint,preprintnumbers,nofootinbib,superscriptaddress]{revtex4-1}

\input{header}

\begin{document}

\preprint{DESY-23-131}

\title{{\Large Investigating the fluxes and physics potential of LHC\\ neutrino experiments}}

\author{Felix Kling}
\email{felix.kling@desy.de}
\affiliation{Deutsches Elektronen-Synchrotron DESY, Notkestr.~85, 22607 Hamburg, Germany}

\author{Toni M\"akel\"a}
\email{toni.makela@ncbj.gov.pl}
\affiliation{National Centre for Nuclear Research, Pasteura 7, Warsaw, 02-093, Poland}
 
\author{Sebastian Trojanowski}
\email{sebastian.trojanowski@ncbj.gov.pl}
\affiliation{Astrocent, Nicolaus Copernicus Astronomical Center Polish Academy of Sciences, ul. Rektorska 4, 00-614, Warsaw, Poland}
\affiliation{National Centre for Nuclear Research, Pasteura 7, Warsaw, 02-093, Poland}

\begin{abstract}
The initiation of a novel neutrino physics program at the Large Hadron Collider (LHC) and the purpose-built Forward Physics Facility (FPF) proposal have motivated studies exploring the discovery potential of these searches. This requires resolving degeneracies between new predictions and uncertainties in modeling neutrino production in the forward kinematic region. The present work investigates a broad selection of existing predictions for the parent hadron spectra at FASER$\nu$ and the FPF to parameterize expected correlations in the neutrino spectra produced in their decays and to determine the highest achievable precision for their observation based on Fisher information. This allows for setting constraints on various physics processes within and beyond the Standard Model, including neutrino non-standard interactions. We also illustrate how combining multiple neutrino observables could lead to experimental confirmation of the enhanced-strangeness scenario proposed to resolve the cosmic-ray muon puzzle already during the ongoing LHC Run~3. 
\end{abstract}

\maketitle 
\clearpage

\section{Introduction} 
\label{sec:introduction}

The subtle role of neutrinos in the Standard Model (SM) constantly motivates measurements of their interactions across a broad energy spectrum, which also remains essential for testing beyond the Standard Model (BSM) scenarios, cf. Refs.~\cite{Formaggio:2012cpf,Proceedings:2019qno,SajjadAthar:2021prg} for reviews. The far-forward region of the Large Hadron Collider (LHC) is particularly suitable for such studies~\cite{DeRujula:1984pg,Winter:1990ry,Vannucci:1993ud,DeRujula:1992sn,Park:2011gh,Feng:2017uoz,Beni:2020yfy}, as it offers a highly-collimated flux of the most energetic neutrinos ever produced in a laboratory setup. A new neutrino physics program has recently been initiated in this region with dedicated FASER~\cite{FASER:2018ceo,FASER:2018bac,FASER:2019dxq,FASER:2020gpr,FASER:2022hcn} and SND@LHC~\cite{SHiP:2020sos,SNDLHC:2022ihg} experiments. Strikingly, this has already led to the first observations of collider neutrinos~\cite{FASER:2023zcr,SNDLHC:2023pun,CERN-FASER-CONF-2023-002}; see also Refs.~\cite{FASER:2021mtu,Arakawa:2022rmp} for earlier analyses and discussion. The initial measurements pave the way for further studies during the ongoing LHC Run~3, and in the future high-luminosity LHC (HL-LHC) era in the proposed purpose-built Forward Physics Facility (FPF)~\cite{Anchordoqui:2021ghd,Feng:2022inv}.

While neutrinos in the SM interact via electroweak gauge bosons, their studies can also indirectly teach us about QCD. This is due to their origin from decays of various mesons produced in hadronic collisions. Due to the uncertainties in modeling the parent hadron production at large pseudorapidities, various theoretical predictions currently differ by as much as an order of magnitude in the expected neutrino charged-current (CC) event rates in the far-forward region of the LHC. Reducing these uncertainties is among the primary goals of the new neutrino experimental program. This will have far-reaching consequences for our understanding of strong interactions, including parton distribution function (PDF) determination and non-perturbative effects, and also broad implications for astroparticle physics and BSM searches, cf. Refs.~\cite{Anchordoqui:2021ghd,Feng:2022inv,Cruz-Martinez:2023sdv,LHCForwardPhysicsWorkingGroup:2016ote}.

The dominant impact of modeling the parent hadron production is also expected to generate notable correlations between neutrino spectra for different flavors and at specific energy ranges. For instance, charm hadron decays determine the forward tau neutrino flux and can contribute substantially to the high-energy part of the electron and muon neutrino spectrum~\cite{Kling:2021gos}. In this study, we propose to utilize these expected correlations to improve the projected constraining power of the ongoing and future neutrino measurements at the LHC. 

To this end, we construct effective parameterization of the far-forward neutrino spectra by interpolating between the leading predictions obtained based on distinct modeling of the hadron production.\footnote{The code developed for this study is available at \href{https://github.com/makelat/forward-nu-flux-fit}{https://github.com/makelat/forward-nu-flux-fit}. The charm-induced incident neutrino spectra used in the analysis were obtained following the repository of the \href{https://github.com/KlingFelix/ForwardCharm}{FPF Working Group 2}. If these files are used together with the code, relevant literature references should be cited, as given in \cref{tab:spectra}.} We combine observations of interactions for different neutrino flavors, energy, and pseudorapidity to determine the expected precision of such analyses using the Hessian-based approach, similar to PDF fits~\cite{Schmidt:2018hvu}.  According to the Cram$\acute{\textrm{e}}$r-Rao bound, this expected precision is given by the
Fisher Information, which can be easily computed~\cite{cramer-rao,cramer-rao2}. Despite existing uncertainties, a multi-channel approach to studying $\nu$-induced events allows for identifying new effects that cannot be easily mimicked by leading SM predictions of the far-forward neutrino spectra or their combinations. This can be used to place strong constraints or discover such phenomena. We illustrate this for an enhanced strangeness production hypothesis with possible groundbreaking implications for cosmic-ray physics~\cite{Anchordoqui:2016oxy,Baur:2019cpv,Anchordoqui:2022fpn} and for BSM-induced neutrino non-standard interactions (NSI) that can also be probed this way at the LHC~\cite{Kling:2020iar,Ismail:2020yqc,Falkowski:2021bkq,Batell:2021snh}.

The paper is organized as follows. In \cref{sec:methodology}, we discuss our modeling, and provide projected bounds on the far-forward neutrino spectra in \cref{sec:spectraconstraints}. \Cref{sec:applications} is devoted to discussing applications of this methodology to constrain enhanced strangeness production and BSM operators describing neutrino NSI. We conclude in \cref{sec:conclusions}. Further details about our statistical analysis are given in \cref{sec:informationgeometry}.

\section{Methodology} 
\label{sec:methodology}

In our analysis, we first obtain a set of neutrino flux predictions to determine the energy and pseudorapidity distribution of far-forward neutrinos at the LHC. The latter distribution can be well described by the radial distribution of events away from the beam collision axis. These predictions are based on different Monte Carlo (MC) generators and other results in the literature, as discussed below. We then define a parameterized flux model, which is constructed from linear combinations of the individual predictions. Using this input, we estimate an expected number of neutrino CC scattering events in existing and proposed on-axis forward neutrino experiments at the LHC. We discuss the necessary ingredients of this analysis in this section. We then estimate how well the LHC neutrino experiments can constrain the flux model on a statistical level and present the results in \cref{sec:spectraconstraints}.

\subsection{Incident Neutrino Fluxes and Spectra\label{sec:spectra}}

Neutrinos that can reach the far-forward detectors of our interest are produced most abundantly near the ATLAS Interaction Point (IP). The meson decays can be either prompt, e.g., for charm mesons, or displaced from the IP, like for charged pions and kaons. In the latter case, the impact of the LHC magnets and infrastructure must be considered in precise modeling. It effectively suppresses neutrino production at distances larger than about $100\,\textrm{m}$ away from the $pp$ collision point. Importantly, for LHC neutrino energies, $E_\nu\sim \textrm{few hundred GeV}$, and the distance between the IP and the detectors, $L \sim \textrm{few hundred meters}$, one expects a negligible impact from neutrino oscillations unless it is enhanced by BSM effects~\cite{FASER:2019dxq}. Hence, the measured neutrino spectra are directly inherited from the parent hadrons.

Various hadrons contribute to the total neutrino flux measured in the far-forward experiments, although the dominant contributions come from charged pions, kaons, D-mesons, and charmed baryons, cf. Ref.~\cite{Kling:2021gos} for detailed discussion. The pion decays dominate the muon neutrino spectrum for energies up to a few hundred GeV, while electron neutrinos with these energies mostly come from kaon decays. Charm contributions might become important at larger energies above TeV and they also determine the tau neutrino flux. Given differences in modeling of the forward hadronic fluxes between charm and light mesons, i.e., pions and kaons, we treat both contributions separately in our analysis. Below, we briefly discuss the MC tools and predictions used in our study, cf. \cref{tab:spectra} for a summary.

\begin{table}
\centering
\begin{tabular}{c|c||c|c}
\hline
\hline
\multicolumn{2}{c||}{Light mesons ($\pi$, $K$)} & \multicolumn{2}{c}{Charm hadrons ($D$, $\Lambda_c$)}\\ 
Name & Refs & Name & Refs\\ 
\hline
\texttt{SIBYLL~2.3d} & \cite{Ahn:2009wx,Ahn:2011wt,Riehn:2015oba,Fedynitch:2018cbl} & \texttt{SIBYLL~2.3d} & \cite{Ahn:2009wx,Ahn:2011wt,Riehn:2015oba,Fedynitch:2018cbl}\\
\texttt{EPOS-LHC} & \cite{Pierog:2013ria} & \texttt{BKRS} & \cite{Buonocore:2023kna}\\
\texttt{DPMJET~3.2019.1} & \cite{Roesler:2000he,Fedynitch:2015kcn} & \texttt{BDGJKR} & \cite{Bai:2020ukz,Bai:2021ira,Bai:2022xad}\\
\texttt{QGSJET~II-04} & \cite{Ostapchenko:2010vb} & \texttt{BKSS} $k_T$ & \cite{Bhattacharya:2023zei}\\
\texttt{Pythia 8.2} (forward) & \cite{Fieg:2023kld} & \texttt{MS} $k_T$ & \cite{Maciula:2022lzk}\\
\hline
\hline
\end{tabular}
\caption{A list of Monte Carlo tools and predictions with references used to obtain far-forward neutrino spectra employed in our study. We treat pions, kaons, and charm hadrons separately in the statistical analysis. See the text for details.
\label{tab:spectra}}
\end{table}

\begin{description}
\item[Light mesons ($\pi$, $K$)] Light meson production in the forward kinematic region of the LHC cannot be described within perturbative QCD (pQCD). Instead, it is typically modeled using hadronic interaction models, many of which were originally designed for cosmic-ray physics. In our analysis, we employ several most commonly used and publicly available MC generators: \texttt{EPOS-LHC}~\cite{Pierog:2013ria}, \texttt{DPMJET 3.2019.1}~\cite{Roesler:2000he,Fedynitch:2015kcn}, \texttt{QGSJET II-04}~\cite{Ostapchenko:2010vb}, and \texttt{SIBYLL~2.3d}~\cite{Ahn:2009wx,Riehn:2015oba}. We follow their implementation in the \texttt{CRMC} package~\cite{CRMC}. We additionally use light meson spectra predictions obtained with a new dedicated forward-physics \texttt{Pythia 8.2} tune~\cite{Fieg:2023kld}. 

Notably, these tools use different approaches to model forward hadron production, and their variation incorporates a variety of underlying physics effects, cf. Refs.~\cite{LHCForwardPhysicsWorkingGroup:2016ote,Albrecht:2021cxw} for reviews. The corresponding predictions form an envelope around the LHCf data on neutral hadron spectra, although there remain sizable variations between them, cf. Refs.~\cite{LHCf:2012mtr,LHCf:2015nel,LHCf:2015rcj,Adriani:2023tyb} for comparison. The first forward muon~\cite{FASER:2023zcr,SNDLHC:2023pun} and electron~\cite{CERN-FASER-CONF-2023-002} neutrino data obtained during the current LHC Run~3 show a broad overall agreement with theoretical predictions that we use, albeit with large statistical uncertainties. 
 
We treat pions and kaons independently in our analysis. To study the robustness of our results, we have performed several numerical tests with a limited set of only three MC generators out of the list of five above and found similar bounds. However, we use the above complete MC generator list in the following. 

\item[Charmed hadrons] Unlike light mesons, charm hadron production can also be described using pQCD. In addition, many of the above generators do not treat forward charm production, or it has not been validated and tuned to LHC data. For this reason, we model the charmed hadron spectra differently in our study. We consider predictions from \texttt{SIBYLL~2.3d}~\cite{Ahn:2011wt,Fedynitch:2018cbl} and, additionally, use several recent results prepared specifically for the far-forward neutrino searches at the LHC. We denote them in the following by acronyms: \texttt{BDGJKR}~\cite{Bai:2020ukz,Bai:2021ira,Bai:2022xad}, \texttt{BKRS}~\cite{Buonocore:2023kna}, \texttt{BKSS} $k_T$~\cite{Bhattacharya:2023zei}, and \texttt{MS} $k_T$~\cite{Maciula:2022lzk}.

Forward charm production in \texttt{SIBYLL} is modeled phenomenologically by replacing the production of a strange pair $s\bar{s}$ by a charm $c\bar{c}$ pair with a small probability determined by fitting to the data~\cite{Ahn:2011wt}. Instead, the remaining predictions employ pQCD calculations of the charm production cross section. The next-to-leading order (NLO) results are used to obtain the \texttt{BKRS} and \texttt{BDGJKR} spectra within the collinear factorization approach. The former calculation uses \texttt{POWHEG}~\cite{Nason:2004rx, Frixione:2007vw, Frixione:2007nw,  Alioli:2010xd} and the NNPDF3.1sx+LHCb set of parton distribution functions (PDFs) with $\alpha_s = 0.118$ at NLO+NLL$_x$ accuracy as input~\cite{Ball:2017otu, Bertone:2018dse}.  The latter results, using the framework of Ref.~\cite{Bai:2021ira}, are obtained with the PROSA FFNS PDF~\cite{Zenaiev:2019ktw} with renormalization and factorization scales proportional to transverse mass set by fitting to the LHCb data.  The \texttt{BDGJKR} predictions include additional Gaussian $k_T$ smearing introduced to mimic the effect of the intrinsic transverse momentum of initial state partons and soft gluon emissions. In contrast, the \texttt{BKSS} $k_T$ and \texttt{MS} $k_T$ model these effects within the hybrid $k_T$ factorization approach~\cite{Dumitru:2005gt,Deak:2009xt}. The Kutak-Sapeta gluon unintegrated PDF (uPDF)~\cite{Kutak:2014wga} is used in this case.

An important effect on the forward charm hadron spectra is related to modeling hadronization and fragmentation. The \texttt{BDGJKR} and \texttt{MS} $k_T$ results are based on applying the Peterson fragmentation function (FF)~\cite{Peterson:1982ak} by assigning a fraction of the momentum of the parent charm quark to the final-state hadron in the partonic center-of-mass frame and laboratory frame, respectively. We note, however, that this calculation neglects the impact of hadronization with beam remnants. Hence, in general, FFs are not expected to be applicable in forward collisions at the LHC, cf. section 6.2.2 in Ref.~\cite{Feng:2022inv} for further discussion. In particular, using them implies that charm hadrons are always less energetic than charm quarks, which reduces the flux of high-energy neutrinos. In the \texttt{MS} $k_T$ case, additional hadronization with beam remnants is also considered via a recombination formalism which is sizeable for $D_0$ and $D^\pm$ mesons but negligible for $D_s$. This effect dominates at high energies and for forward rapidities. On the other hand, \texttt{SIBYLL}, \texttt{BKRS}, and \texttt{BKSS} $k_T$ predictions rely on string fragmentation to include hadronization with beam remnants. The latter two results employ the string fragmentation model implemented in \texttt{Pythia 8.2}~\cite{Sjostrand:2014zea}.
\end{description}

\subsection{Neutrino Flux Parameterization\label{sec:parameterization}}

The forward hadron spectra predictions mentioned above are used to obtain neutrino spectra arising from the decays of the light mesons $\pi^\pm$, $K^\pm$, $K^0_L$, $K^0_S$, and the charmed hadrons $D^\pm$, $D^0$, $\overline{D}^0$, $D^\pm_s$, $\Lambda^\pm_c$. To treat possible variations in the normalization and shape of the neutrino spectra, we take the actual spectra used in our analysis as an interpolation (or extrapolation) between these predictions. For simplicity, we neglect subdominant production modes of neutrinos in hyperon and B-meson decays, as well as secondary production modes in hadronic showers induced in the elements of the LHC infrastructure away from the ATLAS IP. 

To rescale the flux components and to obtain the corresponding binned spectra, we define a model parametrizing the contributions of different predictions in a weighted sum, resulting in a total sample. The parent hadrons are divided into three classes: pions ($\pi$), kaons ($K$), and charmed hadrons ($c$), each with a dedicated weight in the sum. Then with $p \in \{ \pi, K, c \}$, we employ $N_p$ predictions for the number of CC scattering events in the detector in a given energy and radial bin, $G_{n \geq 0}^{(p)}$, by introducing $N_p - 1$ nuisance parameters $\lambda_{i \geq 1}^{(p)}$ to obtain the interpolated prediction with the following expression
\begin{widetext}
\begin{equation}
m =
\sum_{p\in\{ \pi, K, c \}}
\frac{1}{N_p} \left[ G_0^{(p)} \left( 1 - \sum_{i=1}^{N_p-1} \lambda_i^{(p)}
                           \right)
                     + \sum_{i=1}^{N_p-1} G_i^{(p)}
                       \left( 1 + N_p \lambda_i^{(p)} - \sum_{j=1}^{N_p-1} \lambda_j^{(p)}
                       \right)
              \right].
\label{modelExpression}
\end{equation}
\end{widetext}
The model then reduces to the contribution of the $i\geq1$th prediction $G_i$ when 
$\lambda_{i} = 1,\lambda_{j\neq i} = 0$, 
while $\lambda_i = -1~\forall~i$ returns the spectrum of $G_0$. 
Setting $\lambda_i = 0~\forall~i$ yields the average of all predictions, chosen as the baseline for the discussion below. Note that such a setting is not imperative for implementing the model calculation, and choosing the baseline as a general set of parameter values is also possible. In particular, we will discuss the result obtained for the \texttt{SIBYLL} baseline prediction in \cref{sec:strangeness}.

The effective description of the neutrino data obtained this way is characterized by $12$ nuisance parameters, on top of additional free parameters that we introduce when constraining specific new effects discussed in \cref{sec:applications}. While future studies will keep refining the choice of the nuisance parameters in analyses of this kind, the present work is the first quantitative assessment of employing such parameterizations to study LHC neutrinos. These are introduced to relate far-forward neutrino data to fundamental hadronic physics, instead of treating neutrino spectra as fully uncorrelated.

We then perform a likelihood-based analysis and estimate a minimal variance of the model parameters via the Fisher information matrix, as dictated by the Cram$\acute{\textrm{e}}$r-Rao bound~\cite{cramer-rao,cramer-rao2}; see also Refs.~\cite{Cranmer:2014lly,Brehmer:2016nyr,Brehmer:2017lrt} for similar discussions for other LHC data analyses. To this end, our procedure should reproduce the projected most robust bounds to be obtained thanks to the data gathered in considered experimental searches after profiling over nuisance parameters that represent theoretical uncertainties. At the same time, we also comment on expected deviations from this picture in the presence of finite efficiency factors affecting the measurements. The results are, eventually, translated into physically meaningful quantities for their interpretation. We provide more details about a statistical analysis in \cref{sec:informationgeometry}.

In the following, we will focus on the constraints on the combined neutrino and antineutrino spectrum for each flavor, $\nu_\ell+\bar{\nu}_\ell$. We note that the forward LHC detectors have capabilities to disentangle between neutrinos and antineutrinos, especially for $\nu_\mu$. This allows for measuring their spectra separately. We leave the discussion about the potential consequences of such measurements for future studies while we concentrate in this analysis on the dominant impact of meson decays that can be well constrained by the combined spectra.

\subsection{Neutrino Detection\label{sec:detection}}

The collimated flux of high-energy forward neutrinos produced at the LHC can be detected in relatively small experiments that allow for detailed studies of neutrino interactions. We will illustrate the prospects of these searches for a selection of such ongoing and future proposed detectors.

\begin{description}
\item[FASER$\nu$] Focusing first on the current LHC Run~3, we will study the projected capabilities of the FASER$\nu$ emulsion detector~\cite{FASER:2019dxq,FASER:2020gpr}. It consists of tungsten target material layers with a total mass of $1.1$ ton. These are interleaved with emulsion films with the transverse size of $25~\textrm{cm}\times 30~\textrm{cm}$ that store information about the tracks of charged particles produced in neutrino scatterings. High-energy muons produced this way can travel through the entire detector, and their momentum is measured in the FASER spectrometer placed downstream of the emulsion detector. The excellent spatial resolution of emulsion films allows for measuring $\nu_\tau$-induced tau lepton tracks with a few hundred GeV energy and, therefore, study $\nu_\tau$ charged current (CC) interactions on an event-by-event basis.

The expected vertex detection efficiency of FASER$\nu$ is of order $90\%$ for the most energetic neutrinos produced at the LHC, while it decreases to about $(30\%-40\%)$ for $E_\nu\sim 100~\textrm{GeV}$. We implement it following Fig.~9 in Ref.~\cite{FASER:2019dxq}. We additionally employ a geometrical acceptance factor of $80\%$ and lepton identification efficiencies of $86\%$ for muons and $75\%$ for taus following that study. We assume that electrons can be identified with nearly $100\%$ detection efficiency in emulsion due to their expected showering. We note, however, that this identification might become more challenging at lower energies. In particular, in the current analysis, electron neutrino interactions in FASER$\nu$ are studied only above $100~\textrm{GeV}$ energy. We include this effective cut when analyzing FASER$\nu$ prospects for probing the cosmic-ray muon puzzle, as discussed in \cref{sec:strangeness}. Considering all the effects above, we estimate that, e.g., one can identify a CC scattering of the 1 TeV muon neutrino with more than $60\%$ efficiency in FASER$\nu$. In this analysis, we use $5$ energy bins per decade in the likelihood analysis, which can reproduce expected $30\%$ neutrino energy resolution in this detector~\cite{FASER:2019dxq}. We assume $\mathcal{L} = 150~\textrm{fb}^{-1}$ of integrated luminosity in LHC Run~3.

\item[FASER$\nu$2] The emulsion detector technology has also been proposed for the FASER$\nu$2 detector in the FPF. The assumed transverse size of $40~\textrm{cm}\times 40~\textrm{cm}$ and total tungsten mass of $20$ tons, as well as larger integrated luminosity in the HL-LHC era, $\mathcal{L} = 3~\textrm{ab}^{-1}$, result in a significantly increased expected neutrino event statistics in this detector, up to 1M muon neutrino CC scatterings~\cite{Anchordoqui:2021ghd,Feng:2022inv}. The larger detector size of FASER$\nu$2 permits better event containment than in FASER$\nu$. This results in an expected improvement in energy resolution. We, therefore, employ $10$ bins per decade of the incident neutrino energy in this case. Similarly to FASER$\nu$, the neutrino detection efficiency in FASER$\nu$2 will be flavor-dependent. Given the lack of detailed efficiency studies for FASER$\nu$2, we present the results below assuming $100\%$ efficiency. However, we also comment on the impact of employing efficiency cuts similar to those discussed above for the currently operating FASER$\nu$ detector. 

\item[FLArE] We also present the results for the proposed FLArE detector~\cite{Batell:2021blf,Anchordoqui:2021ghd,Feng:2022inv} employing liquid argon (LAr) time-projection chamber (TPC) technology. FLArE will offer improved calorimetric capabilities and dynamical information about events to disentangle neutrino-induced signals from muon backgrounds. The outgoing muons from neutrino interactions can be measured with a dedicated muon tagger and with the help of the FASER2 spectrometer. Studying tau neutrinos might be more challenging in this case due to the expected lower spatial resolution of LArTPCs than in emulsion detectors. However, $\nu_\tau$-induced events can still be searched for as fluctuations over the expected backgrounds from other neutrino flavors. In the following, we assume $1~\textrm{m}\times 1~\textrm{m}$ transverse area and 10-ton fiducial mass of the LAr target in FLArE, and the integrated luminosity of $\mathcal{L} = 3~\textrm{ab}^{-1}$. We take $100\%$ efficiency for neutrino detection in FLArE while commenting on the case with a decreased $50\%$ efficiency. 
\end{description}

All the detectors discussed above are centered around the beam-collision axis. Importantly, off-axis far-forward detectors have also been proposed, namely the SND@LHC~\cite{SHiP:2020sos,SNDLHC:2022ihg} and AdvSND~\cite{Anchordoqui:2021ghd,Feng:2022inv} experiments for the ongoing LHC Run~3 period and the HL-LHC era, respectively. These extend pseudorapidity coverage of far-forward searches at the LHC toward lower values of $\eta$. In the following, we focus on the on-axis experiments and present representative results obtained for the ongoing measurements in FASER$\nu$ and the proposed FASER$\nu$2 and FLArE searches. We note, however, that additional data gathered off-axis may further improve the projected constraints discussed below.

When modeling neutrino interactions in the detectors of our interest, we convolute the neutrino flux with the interaction cross-sections predicted by \texttt{GENIE}~\cite{Andreopoulos:2009rq} as obtained in Ref.~\cite{FASER:2019dxq}. These results are based on a Bodek-Yang model used to describe deep inelastic scattering (DIS) events~\cite{Yang:1998zb,Bodek:2002vp}. The alternative NNFS$\nu$ approach has been recently discussed in Ref.~\cite{Candido:2023utz}, which generally agrees with the Bodek-Yang model at TeV-scale energies, cf. also Refs.~\cite{Xie:2023suk,Jeong:2023hwe} for other recent analyses. However, uncertainties in the predicted scattering cross section up to a few percent for $E_\nu\sim \textrm{TeV}$ have been reported that are driven by PDF uncertainties~\cite{Candido:2023utz}. This is not expected to significantly affect the interpretation of the results presented below for the ongoing FASER$\nu$ measurements. On the other hand, improved sensitivity of the FPF experiments will allow us to reach the level of precision where PDF uncertainties are anticipated to become important. In fact, by using additional kinematic variables, the FPF is expected to constrain PDFs, especially for strange quarks~\cite{Anchordoqui:2021ghd,Feng:2022inv}. The proposed Electron-Ion Collider (EIC) will further improve relevant bounds on up and down quark PDFs~\cite{Accardi:2012qut}. The corresponding uncertainties should then be reduced during the FPF data-taking period. In the following, we focus on the dominant uncertainties affecting neutrino fluxes and spectra in the far-forward kinematic region of the LHC related to the differences in parent hadron spectra predictions. We leave the discussion of a joint fit considering both production and interaction rate uncertainties for the future.

\section{Neutrino spectra and projected constraints\label{sec:spectraconstraints}}

\begin{figure*}[t]
\centering
\includegraphics[width=0.9\textwidth]{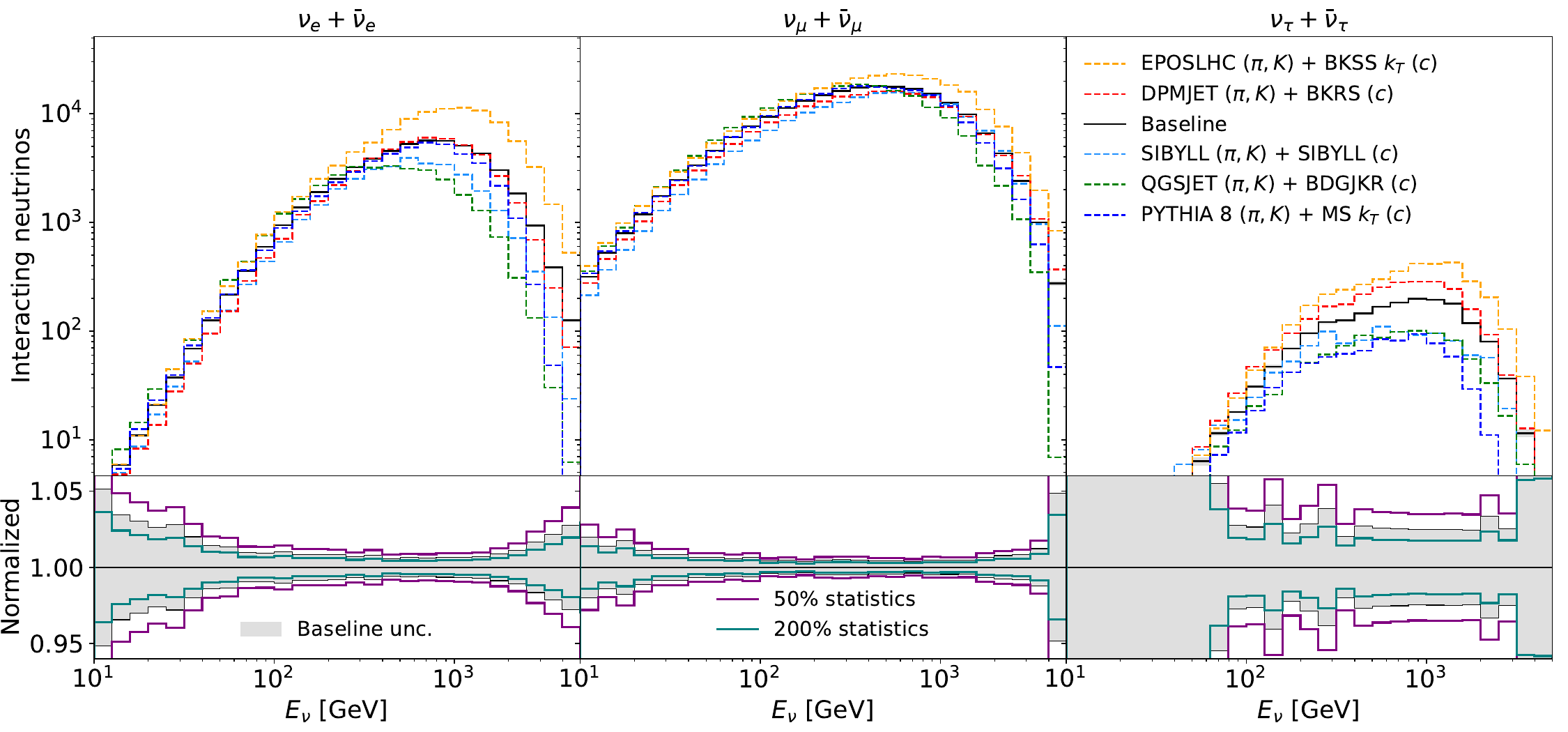}
\caption{In the upper panels, the colorful histograms correspond to different predictions of the combined energy distributions of neutrinos and antineutrinos interacting via CC scatterings in FLArE, as indicated in the plot. The left (central, right) panel corresponds to the electron (muon, tau) neutrinos. An average of the predictions employed in the analysis gives the baseline spectrum shown with a black solid line. The bottom panels illustrate the expected Cramer-Rao uncertainty bands ($1\sigma$) on the baseline spectrum as gray-shaded regions. The robustness of the obtained uncertainties against varying event statistics is shown with purple and green histograms, where the number of events is changed up and down by a factor of two.}
\label{spectra_stat_var}
\end{figure*}

In the upper panels of \cref{spectra_stat_var}, we illustrate single-differential neutrino energy distributions for CC scattering events in the FLArE detector using several combinations of the abovementioned MC predictions for parent meson spectra. We present the results for all three neutrino flavors. We denote different predictions by $\textrm{p}_1 + \textrm{p}_2$ in the plots, where $\textrm{p}_i$ stands for the prediction name, and $i=1$ and $2$ corresponds to light and charm hadron spectra, respectively. In each case, the plots show the combined neutrino and antineutrino spectra.

As can be seen, various predictions agree remarkably well for the electron and muon neutrinos with energies up to $E_\nu\sim 300~\textrm{GeV}$. In this energy regime, an observed discrepancy between different MC results is about a factor of $2$. This reflects a relatively better understanding of light meson spectra production in the far-forward region of the LHC, and these mesons dominate the $\nu_e$ and $\nu_\mu$ fluxes up to a few hundred GeV of energy. Instead, the larger the neutrino energy becomes, the uncertainties grow both for light mesons and especially for the possible charm hadron contributions. The latter also determine the $\nu_\tau$ flux predictions over the entire energy range. The charm-induced spectra currently show an order-of-magnitude discrepancy between various predictions.

Focusing on the tau neutrino spectrum plot, we find that the lack of beam remnant induced effects in hadronization, e.g. the beam drag effect in modeling the $D_s$-meson production, suppresses the high-energy part of charm-induced neutrino spectra. This is evident when comparing the \texttt{BDGJKR} and \texttt{MS} $k_T$ predictions with the \texttt{BKRS} and \texttt{BKSS} $k_T$ results. We note that even though the high-energy part of the BKSS $\textrm{k}_\textrm{T}$ spectrum is suppressed by considering gluon saturation, this prediction remains the most optimistic in terms of the expected number of $\nu$-induced events in the detector. The difference between this prediction and the least optimistic \texttt{MS} $k_T$ result is the largest for the most energetic tau neutrinos with $E_{\nu_\tau}\sim \textrm{few TeV}$. Furthermore, we have verified that the uncertainties in the charm predictions also partially propagate to the high-energy part of $\nu_e$ and $\nu_\mu$ spectra, adding to uncertainties in determining light meson spectra. 

We also show in the plots the baseline model prediction obtained as an average of all the considered predictions, assuming equal weights. In the bottom panels in \cref{spectra_stat_var}, we assume that the baseline prediction correctly describes the data to be gathered in the FPF. The gray-shaded regions illustrate the projected statistical precision with which our flux model can be constrained at $1\sigma$ level; see \cref{sec:informationgeometry} for details of the statistical analysis. 

The uncertainty bands found this way illustrate excellent precision in constraining the neutrino spectra in the FPF experiments. This is especially evident for muon neutrinos with energies $100~\textrm{GeV}\lesssim E_{\nu_\mu}\lesssim 1~\textrm{TeV}$, as shown in the bottom central panel in the figure. Due to the largest expected event statistics, the projected bounds, in this case, are at the percent level. This translates into a narrow gray uncertainty band over the baseline neutrino spectrum in the central upper panel, which is barely visible in the plot. In particular, the FPF data will allow for differentiating between the baseline hypothesis and specific MC results presented in \cref{spectra_stat_var} with high precision. 

Due to reduced event statistics, the uncertainty bands grow at the spectrum's low and high energy tails. The high-energy neutrinos with $E_\nu\gtrsim \textrm{a few} \times \textrm{TeV}$ are more rarely produced at the LHC. Instead, low-energy neutrinos with $E_\nu\lesssim 10~\textrm{GeV}$ are produced more isotropically and often miss far-forward experiments. However, we find the projected uncertainty to be of order several percent between these two regimes. This remains at the level of PDF uncertainties affecting the neutrino DIS cross-section predictions, as discussed above. This happens also for the electron neutrinos, for which the expected number of events is only a factor of a few lower than for $\nu_\mu$s. We show the electron neutrino uncertainty bands in the bottom left panel.

The bottom right panel illustrates the results for the tau neutrinos. In this case, the projected uncertainties are larger but, remarkably, also stay below $5\%$ for $100~\textrm{GeV}\lesssim E_{\nu_\tau}\lesssim 3~\textrm{TeV}$. At first, this result might seem odd, given significantly lower event statistics of $\nu_\tau$-induced events than for the other neutrino flavor. However, we note that the analysis for the tau neutrinos implicitly concerns the results obtained for both $\nu_e$ and $\nu_\mu$. This is because the spectra of these neutrinos are also affected by the forward charm production, especially in their high-energy tails. Possible enhanced production of charm hadrons is then strongly constrained in this energy regime by the electron and muon neutrino data, which then translates into stronger bounds on $\nu_\tau$. Instead, in the low-energy part of the spectrum, below $100~\textrm{GeV}$, both the tau neutrino flux is decreased, and the correlation with the electron and muon neutrino spectra is lost. As a result, the constraining power for $\nu_\tau$ in this energy regime is significantly weaker. 

We have also verified numerically that the expected uncertainty bands on the $\nu_\tau$ energy spectrum depend only mildly on the choice of the baseline spectrum. For instance, after switching to the baseline spectrum defined as \texttt{DPMJET}($\pi$,$K$) + \texttt{BKRS}($c$) shown in red in \cref{spectra_stat_var}, one finds reduced uncertainties, by up to a factor of $2$, in some of the low-energy bins for $E_{\nu_\tau}\lesssim 100~\textrm{GeV}$. The improvement in high-energy bins is, however, much smaller, even though the new baseline spectrum predicts a larger number of $\nu_\tau$-induced events up to $E_{\nu_\tau}\sim \textrm{TeV}$. This additionally illustrates that the high-energy tail of the tau neutrino spectrum is not only sensitive to the $\nu_\tau$ spectrum, but the charm contribution to the spectra of other neutrino flavors strongly constrains it too. The latter constraining power is not significantly affected by changing the baseline spectrum. This is because \texttt{DPMJET} predictions accidentally lie close to the average spectra for $\nu_e$ and $\nu_\mu$ over the entire energy range, as can be seen by comparing red and black histograms in the left and central upper panels of \cref{spectra_stat_var}.

In \cref{spectra_stat_var}, we also illustrate the expected uncertainty bands for each neutrino flavor that assume only $50\%$ of event statistics. We show this with purple histograms in the bottom panels. As discussed above, this could correspond to a more realistic treatment of the neutrino detection efficiency factors in FLArE. Importantly, as can be seen, this has only a mild impact on the expected constraining power of this experiment. Similarly, we present the expected results for increased event statistics up to $200\%$ of events with green histograms in the bottom panels. This could be due to increasing the fiducial volume of the detector. Again, the predicted impact on the neutrino spectrum uncertainty bands is relatively small. Hence, small variations in efficiency factors or detector sizes in the FPF are not expected to affect the neutrino physics program significantly.

\begin{figure*}[t]
\centering
\includegraphics[width=0.9\textwidth]{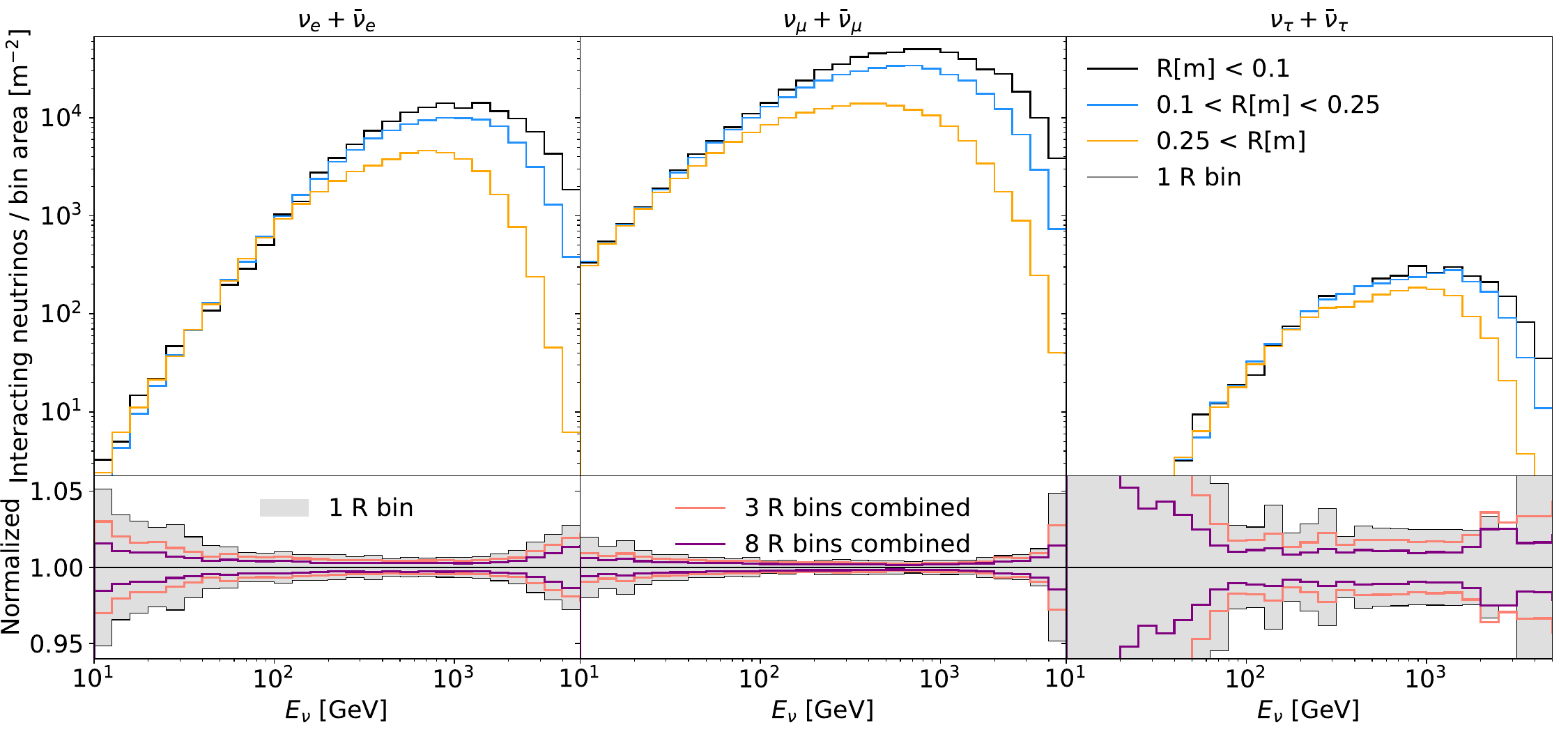}
\caption{The upper panel illustrates the combined neutrino and antineutrino CC event scattering rates in FLArE, using the same baseline spectrum as Fig.~\ref{spectra_stat_var}. The results are shown for each neutrino flavor in three radial bins, as indicated in the plot. The spectra are divided by the corresponding bin area. The lower panel indicates the improvement in uncertainty obtained by combining the information from three (red) or eight (purple) radial bins.}
\label{radial_summary}
\end{figure*}

However, adding spatial information about events can improve the neutrino spectrum uncertainty bands. This allows for constraining double-differential neutrino production cross section in the far-forward region of the LHC, which takes into account additional information about the pseudorapidity distribution on top of the previously discussed energy distribution. We illustrate this in \cref{radial_summary}, in which the spatial distribution of neutrino scattering events in FLArE is considered by virtue of radial bins, using the same baseline spectrum as considered in \cref{spectra_stat_var}. In the upper panels, we show the neutrino interaction spectrum in three radial bins defined as $R<0.1~\textrm{m}$, $0.1~\textrm{m}<R<0.25~\textrm{m}$, and $R>0.25~\textrm{m}$, where $R$ is the radial distance away from the beam collision axis. The detector is assumed to be centered around the beam collision axis ($R=0$), and the last radial bin extends to the edges of the detector transverse size defined by the square of size $1~\textrm{m}\times 1~\textrm{m}$. The spectra are normalized to the bin area to illustrate better the concentration of neutrino-induced events around the beam collision axis.\footnote{In the analysis below, we also use radial bins for the other experiments that are defined as follows. For FASER$\nu$, with the smallest transverse size, we use $R<0.06~\textrm{m}$, $0.06~\textrm{m}<R<0.13\textrm{m}$, and $R>0.13~\textrm{m}$ up to the edge of the detector. In the case of FASER$\nu$2, we define the bins differently: $R<0.1~\textrm{m}$, $0.1~\textrm{m}<R<0.2~\textrm{m}$, and $R>0.2~\textrm{m}$ up to the edge of the detector.}

As shown with solid black lines in the upper panels, the central parts of the detector ($R<0.1~\textrm{m}$) can constrain well the most uncertain high-energy parts of the neutrino spectra. Instead, the outermost radial bin in this energy regime is characterized by more than an order of magnitude lower neutrino flux per unit area, as shown with yellow solid lines. This is, however, compensated by a larger area of this radial bin when counting the total number of events. Hence, each radial bin has similar constraining power in our analysis in the high-energy tails of the distributions. Instead, neutrinos with lower energies, below a few hundred GeV, are dominantly constrained by the data gathered in the parts of the detector with a larger total transverse area. This is understood as their parent mesons are often less energetic and less forward-focused after production at the LHC.

Considering this spatial information further improves the FPF detectors' constraining power. We illustrate this in the bottom panels of \cref{radial_summary}. In the plots, gray-shaded regions correspond to the previously discussed results with only one radial bin. In this case, only a single-differential distribution in the energy of the neutrino production cross section is used to constrain neutrino spectra. Instead, red and purple lines in the plots show the results obtained for three or eight radial bins. As can be seen, adding spatial information reduces the uncertainties to the sub-percent level for the muon neutrinos with $100~\textrm{GeV}\lesssim E_{\nu_\mu}\lesssim \textrm{TeV}$. A similar reduction is observed for the electron neutrinos. The improvement by up to a factor of a few in the expected uncertainty band is also found in the low- and high-energy tails of the respective neutrino spectra. Increasing the number of radial bins further does not substantially improve the uncertainty bands. This is due to reduced event statistics in each of the bins observed in this case. 

The baseline spectrum uncertainty for $\nu_\tau$s is, similarly, reduced over the entire energy range by using spatial information. In particular, the low-energy tail of the spectrum obtained for $E_{\nu_\tau}\sim \textrm{a few tens of GeV}$ can now be better constrained. Charm-induced neutrinos are characterized by a noticeably different pseudorapidity distribution than those produced in decays of light mesons. The latter tend to be more collimated around the beam collision axis, as dictated by their characteristic transverse momentum, $p_T\sim m/p$, where $m$ is the hadron mass and $p$ is its total momentum. Therefore, including information about the double-differential distribution allows for better disentangling charm-induced excess of $\nu_e$ and $\nu_\mu$ scattering events over the dominant events associated with the neutrino production in light meson decays. The improved charm constraining power also reduces uncertainty bands on the $\nu_\tau$ spectrum.

\begin{figure*}[t]
\centering
\includegraphics[width=0.9\textwidth]{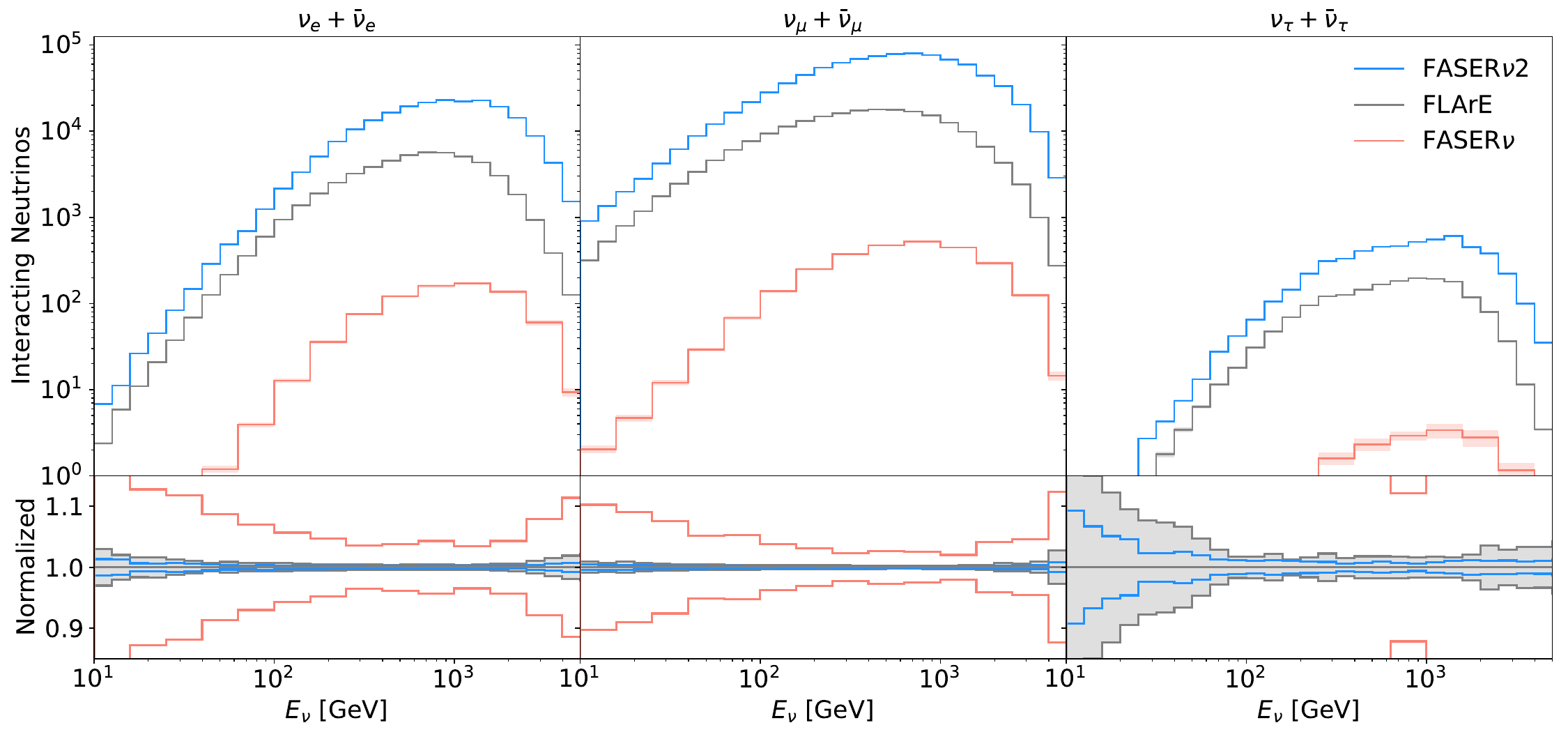}
\caption{Similar to \cref{spectra_stat_var}, but a comparison of the baseline neutrino CC scattering interaction rates obtained for FLArE, FASER$\nu$2, and FASER$\nu$ are shown, assuming luminosities of $150\mathrm{fb}^{-1}$ for FASER$\nu$ and $3\mathrm{ab}^{-1}$ for the remainder. In the bottom panels, relevant uncertainty bands are shown.}
\label{experiment_comparison}
\end{figure*}

In \cref{experiment_comparison}, we show a comparison between the baseline neutrino spectra and uncertainty bands obtained for FLArE and FASER$\nu$2 in the FPF and the currently operating FASER$\nu$ detector. As can be seen, the FPF experiments will offer more than two orders of magnitude larger neutrino event statistics than FASER$\nu$. The highest number of events is expected for FASER$\nu$2, which, according to the current design, has a larger target mass by a factor of two than FLArE. Additional improvement comes from an increased tungsten density with respect to LAr. This allows for concentrating the target mass better around the beam collision axis, where high-energy neutrino flux is collimated. Because of the larger transverse size of FLArE, a peak of the expected neutrino spectrum in this detector is slightly shifted toward lower energies when compared to emulsion detectors. 

The increased event statistics in the FPF detectors translate into significantly narrower uncertainty bands than for FASER$\nu$, as shown in the bottom panels. These have been obtained assuming $3$ radial bins for each detector. The relevant ranges of $R$ have been changed for each detector, depending on its transverse size. Notably, the ongoing measurements in FASER$\nu$ will be able to constrain the electron and muon neutrino spectra with $\mathcal{O}(10\%)$ precision for the energy between a few hundred GeV and several TeV. However, the uncertainties in determining the tau neutrino flux will remain much larger. The FPF detectors are needed to reduce them to a few percent level.

\section{Physics Applications} 
\label{sec:applications}

As discussed above, detailed information about the neutrino flavor, energy spectrum, and the spatial distribution of events in the detector will allow one to differentiate between various predictions. It can also be used to constrain other effects. Employing complete information about events allows for better identification of the unique impact of such phenomena on the far-forward neutrino data. We illustrate this below for two sample effects. One is related to proposed enhanced strangeness production in hadronic collisions at large energies and pseudorapidities. The other effect concerns potential NSI contributions to neutrino event rates in the far-forward neutrino experiments at the LHC.

\subsection{Enhanced Strangeness}
\label{sec:strangeness}

Far-forward searches at the LHC are naturally connected to ultra-high energy cosmic-ray (UHECR) physics. This is due to the sensitivity of both physics programs to high-energy hadronic collisions and the importance of large pseudorapidity regimes of such interactions. We have already shown how LHC data can help differentiate between available MC generators that are also routinely used in modeling cosmic ray (CR) air showers to tune them better in the future. Here, we focus on the expected impact of these searches on explaining anomalies in cosmic-ray data.

A striking example of such anomaly is the so-called \textsl{muon puzzle} first observed in the Pierre Auger Observatory data~\cite{PierreAuger:2014ucz,PierreAuger:2016nfk,PierreAuger:2021qsd}. Other experimental collaborations subsequently confirmed it, and the anomaly is currently considered to have a combined statistical significance of $8\sigma$, cf. Ref.~\cite{Albrecht:2021cxw} for review. The anomaly is related to an apparent enhancement in muon rates at the level of a few tens of percent in hadronic components of CR-induced showers. This corresponds to high energies of the incident CR starting at $E\sim 10^8~\textrm{GeV}$, which translates into $\sqrt{s}\simeq \sqrt{2\,E\,m_p}\simeq 14~\textrm{TeV}$ in the CM frame of the $pp$ collision between the CR and proton in oxygen or nitrogen nuclei in the atmosphere. Notably, this is the energy scale characteristic for the LHC. The discrepancy between the observed and predicted muon rates grows higher with increasing energy. It has been shown that the dominant explanation of the anomaly is likely due to a reduced transfer of energy from a hadronic to an electromagnetic component of the shower, e.g., by suppressing the neutral pion production or decay rate in atmospheric air showers~\cite{Allen:2013hfa}.

Among the models proposed to accommodate such an effect, particularly important is the \textsl{enhanced strangeness} hypothesis, in which suppressed pion to kaon production ratio in the final state of high-energy $pp$ collisions is assumed, cf. Refs.~\cite{Anchordoqui:2016oxy,Baur:2019cpv} for possible underlying mechanisms. In a simple phenomenological approach, this can be achieved by introducing a finite swapping probability that turns a fraction of pions into kaons. A detailed study of this effect has been performed in Ref.~\cite{Anchordoqui:2022fpn}. It has been shown that the relevant $\pi\to K$ swapping fraction $f_s$ at the level of a few tens of percent can explain the anomaly. To this end, and to be reconciled with other experimental data, the swapping probability should primarily affect high-energy collisions in the large pseudorapidity regime. Interestingly, hints of enhanced strangeness production have also been found in the mid-rapidity region in the ALICE data~\cite{ALICE:2016fzo}.

In the following, we will analyze a simple phenomenological model introduced in Ref.~\cite{Anchordoqui:2022fpn}. In this case, in the presence of the non-zero $f_s$ parameter, the number of neutrinos produced from pion decays in the forward region of the LHC is reduced by a common energy-independent factor, $N_{\pi\to\nu}\to (1-f_s)\,N_{\pi\to\nu}$. Simultaneously, the number of neutrinos produced in kaon decays is increased as $N_{K\to\nu} \to (1+6.6\, f_s) N_{K\to\nu}$. Here, a numerical factor of $6.6$ is related to a relative difference in the pion and kaon production rates at large pseudorapidities at the LHC. It has been determined numerically to reproduce best a complete treatment of the model, in which individual pions are changed into kaons in simulations of the forward neutrino spectra. The difference in the production rates of both mesons is due to their different masses and quark compositions. Additional effects considered in these simulations are due to finite kaon lifetimes and the change of $\pi^0$ into $K^0_{S,L}$. In the latter case, the neutrino can only be produced after the swapping, while the initial neutral pion would typically decay into two photons. Assuming \texttt{SIBYLL} as a baseline MC generator, it has been shown that introducing such a universal swapping fraction $f_s$ for collisions characterized by projectile energies above PeV and pseudorapidities $|\eta|>4$ in the CM frame in CR air shower simulations allows for fitting the muon data. This requires $f_s$ to lie between about $0.3$ and $0.8$, where larger values are favored when the increasing primary energy is considered.

\begin{figure*}[t]
\centering
\includegraphics[width=0.66\textwidth,trim={0mm 0mm 120mm 0mm},clip]{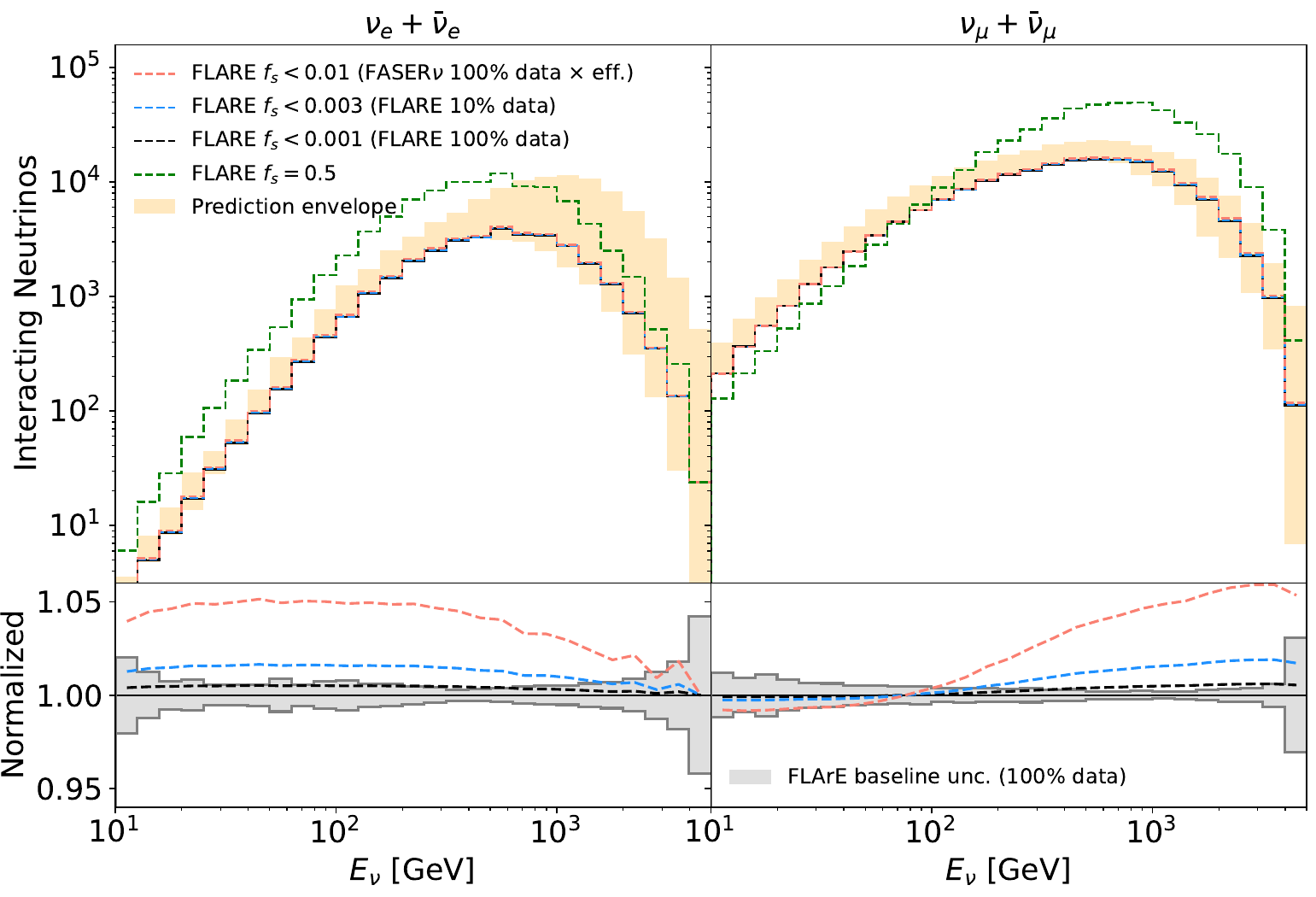}
\includegraphics[width=0.323\textwidth,trim={0mm 0mm 0mm 0mm},clip]{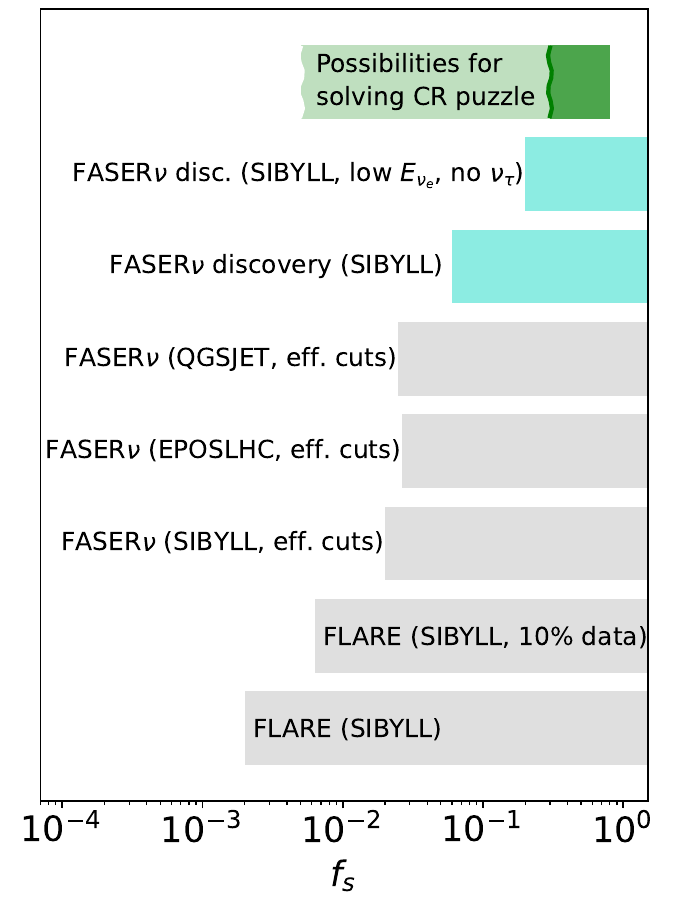}
\caption{\textit{Left:} The top panels show the electron (left) and muon (right) neutrino CC event scattering rates in FLArE obtained using \texttt{SIBYLL} as the baseline MC generator and with three radial bins. The solid black histograms correspond to $f_s = 0$, while the dashed orange (blue, black) ones to $f_s = 0.01$, $0.003$, $0.001$. The latter remain barely distinguishable from the $f_s=0$ baseline in the plots. These values of $f_s$ roughly correspond to $1\sigma$ exclusion bounds obtained for FASER$\nu$ and FLARE with 10\% or 100\% of total data. For FASER$\nu$, the efficiency factors arising from geometry, energy dependence, and charged lepton identification have been applied. The green histograms represent the $f_s=0.5$ case, for which the cosmic-ray muon puzzle can be solved. The variations in the neutrino event rate due to different MC predictions from \cref{spectra_stat_var} are shown with yellow-shaded bands. The bottom panels zoom in on the uncertainty bands on the neutrino spectrum, shown as gray-shaded bands similar to \cref{spectra_stat_var}. Expected deviations from the $f_s=0$ case are also shown as colorful lines that correspond to the aforementioned exclusion bounds from FASER$\nu$ and FLArE.
\textit{Right:} The $2\sigma$ constrained values (gray) for $f_s$ obtained using FLARE and FASER$\nu$, also demonstrating the effect of choosing different predictions as the baseline for the latter. These are compared to less constraining values obtained for the discovery potential at FASER$\nu$ (turquoise), with and without the information on tau neutrinos and high-energy contributions to the $\nu_e$ spectrum. Notably, all of the predicted constraints cover the $0.3 < f_s < 0.8$ region shown in dark green, i.e., the values of $f_s$ favored by the enhanced strangeness solution to the CR muon puzzle. The light green band extending to lower values of $f_s\sim 0.005$ is added to indicate that the effect might manifest in a more subtle way in $pp$ collisions at the LHC.}
\label{enhanced_strangeness}
\end{figure*}

Such effects can be particularly prominent in the forward LHC neutrino data if they change $\nu$ interaction rates in kinematic regions less affected by variations in MC predictions. We illustrate this for the enhanced strangeness effect in the upper left panel of \cref{enhanced_strangeness} with two plots obtained for electron and muon neutrinos. In the plots, we present green histograms representing the expected neutrino CC event scattering rate in the FLArE detector obtained for \texttt{SIBYLL} and $f_s = 0.5$. This should be compared with black solid lines in the plots representing the baseline scenario obtained for $f_s = 0$. As can be seen, the enhanced strangeness production, in this case, would manifestly increase the electron neutrino event rates over the entire energy range, especially for $E_{\nu_e}\lesssim 1~\textrm{TeV}$. This is due to the dominant $\nu_e$ production mode in kaon decays. A similar enhancement is predicted for muon neutrinos above $100~\textrm{GeV}$. Instead, for lower energies, one expects a decrease in the $\nu_\mu$ event statistics, albeit this is a less significant effect driven by a reduced number of forward-going pions. Applying a non-zero swapping probability does not affect the tau neutrino spectrum. A combined impact of these modifications of the neutrino spectra measured in the far-forward region of the LHC provides a strong signature of this effect, which cannot be easily reproduced by changing and interpolating between various MC predictions in our analysis. To illustrate this, we have added yellow-shaded prediction envelopes in the plots around the baseline distributions that correspond to various MC results shown in \cref{spectra_stat_var}.

We first note that essential bounds on the $f_s$ parameter will be obtained thanks to the data gathered in FASER$\nu$ during the ongoing LHC Run~3. Using the procedure outlined above, we have found that already within the next few years, FASER$\nu$ will be able to constrain the enhanced strangeness hypothesis up to the level of $f_s\simeq 0.013$ ($1\sigma$) assuming \texttt{SIBYLL} as a baseline (measured) neutrino spectrum. These results only mildly depend on the precise choice of the baseline spectrum. In particular, we have also verified this for the spectra generated with the \texttt{EPOS-LHC} and \texttt{QGSJET} MC tools and found similar expected bounds at the level of $f_s \simeq 0.013$ and $0.012$, respectively. As discussed in Ref.~\cite{Sciutto:2019pqs}, these MC generators predict either smaller or larger enhancement effects in the CR shower data. Notably, regardless of the precise choice of the generator, the constraining power of FASER$\nu$ significantly exceeds the preferred value of $f_s\sim (0.3-0.8)$ obtained by fitting the UHECR data.

This motivates studying potential discovery prospects in FASER$\nu$. We have tested them assuming that the neutrino data gathered in FASER$\nu$ will correspond to \texttt{SIBYLL} predictions enhanced by an additional impact of the non-zero $f_s$ parameter. We find in this case that the unique features of this scenario differ from other SM predictions sufficiently strongly to allow for excluding the $f_s = 0$ hypothesis at the $5\sigma$ level for the swapping fraction $f_s = 0.06$ or so. We recall that this result has been obtained by considering realistic FASER$\nu$ efficiency factors, as discussed in \cref{sec:detection}. 

To obtain even more baseline-independent results, we similarly study the discovery prospects for FASER$\nu$ focusing only on the muon neutrino data and electron neutrinos with energies in the range $100~\textrm{GeV}\lesssim E_{\nu_e}\lesssim 300~\textrm{GeV}$. This excludes high-energy electron neutrinos and the tau neutrino data that are currently subject to the largest theoretical uncertainties based on various MC predictions, cf. \cref{spectra_stat_var} and yellow-shaded bands in the upper left panels of \cref{enhanced_strangeness}. After limiting the dataset for the enhanced strangeness analysis this way, we still find good discovery prospects in FASER$\nu$. The $f_s= 0$ hypothesis will be then excluded at $5\sigma$ for $f_s \gtrsim 0.2$. This is driven by the low-energy part of the $\nu_e$ spectrum, in which significant deviations from all the MC predictions are expected for $f_s$ of order tens of percent. The capabilities of FASER$\nu$ in probing this effect will be further enhanced by combining the data gathered by this detector and the SND@LHC experiment. We conclude that the ongoing far-forward neutrino physics program at the LHC will be able to decisively test benchmark models predicting a few tens of percent pion to kaon swapping fractions in forward collisions at the relevant energy and probe this solution to the CR muon puzzle.

While LHC Run~3 searches will already place strong constraints on this scenario, it is also possible that the swapping probability might not be a constant factor. In particular, it can depend on the mass number of colliding nuclei and become more substantial for increasing $A$, while it could be less pronounced in $pp$ collisions~\cite{Anchordoqui:2016oxy,Anchordoqui:2022fpn}. In addition, the impact of energy and pseudorapidity dependence of $f_s$ on the CR data has recently been studied in Ref.~\cite{Sciutto:2023zuz}. It has been shown that introducing such dependence can, e.g., allow for solving the puzzle for the linearly increasing $f_s$ parameter with growing energy. This would predict smaller values of $f_s^{\textrm{(LHC)}}$ at LHC energies, while the maximum value $f_s^{\textrm{(max)}}$ would still be large and substantially modify the kaon production rate at higher energies. In the example discussed therein, one can estimate $f_s^{\textrm{(LHC)}}\sim 0.005$ if $f_s^{\textrm{(max)}}\sim 0.5$ is assumed. The muon puzzle can still be solved in this case. It is then possible that only a more subtle impact of the enhanced strangeness scenario could be seen in $pp$ collisions at the LHC.

Going beyond a few percent precision might be then crucial to probe this scenario in the far-forward LHC searches. This will be possible with the proposed FPF experiments. In the bottom left panels of \cref{enhanced_strangeness}, we show in gray the expected uncertainty bands on the electron and muon neutrino spectra in FLArE obtained similarly to \cref{spectra_stat_var}. On top of this, we show the predicted deviations for the pion-to-kaon swapping probability of $1.3\%$, $0.43\%$, and $0.14\%$. These correspond to the FASER$\nu$ $1\sigma$ exclusion bound discussed above and to FLArE constraints obtained with either $10\%$ of data or the full dataset. As can be seen, within the first one to two years of data taking, FLArE will surpass the ongoing LHC searches by a factor of a few in probing the $f_s$ parameter. The improvement by about an order of magnitude in $f_s$ is expected after the entire HL-LHC era such that sub-percent values of this parameter will be tested. 

We summarize the expected bounds on $f_s$ in the right panel of \cref{enhanced_strangeness}. In the plot, we indicate with a dark green color the preferred range of values of the $f_s$ parameter that could explain the CR muon puzzle. We put it between $0.3$ and $0.8$ following the results present in Ref.~\cite{Anchordoqui:2022fpn} for concreteness. We also show in the plot an extended light green band towards lower values of $f_s\sim 0.005$, which refers to the possible smaller magnitude of this effect in $pp$ collisions at the LHC. On top of this, we show in turquoise the $f_s$ ranges that can lead to discovery in the ongoing FASER$\nu$ searches, based on either the full neutrino data or a limited dataset to $\nu_\mu$ and low-energy electron neutrinos. We also present in the figure a set of gray-shaded exclusion bands at $2\sigma$ obtained for FASER$\nu$ with three different baseline MC generators and for FLArE with the entire or limited data sets, as discussed above. The proposed FLArE experiment will probe this scenario up to $\mathcal{O}(0.1\%)$ level in $f_s$, below which barely any effect on the $pp$ final-state meson distribution is expected.

\subsection{Neutrino Charged Current Non-Standard Interactions}
\label{sec:NSI}

One of the major developments of the far-forward neutrino physics program at the LHC is the possibility of studying CC interactions of the tau neutrinos at the TeV energy scale on an event-by-event basis. This is thanks to the exceptional capabilities of the currently operating emulsion detectors that could be further improved in the future in the FPF experiments. Below, we discuss how these searches can help to constrain possible new physics contributions to high-energy neutrino interactions, cf. Refs.~\cite{Bahraminasr:2020ssz,Kling:2020iar,Bakhti:2020szu,Jodlowski:2020vhr,Ismail:2020yqc,Falkowski:2021bkq,Ismail:2021dyp,Ansarifard:2021elw,Kelly:2021mcd,Cheung:2021tmx,Ansarifard:2021dju,Cheung:2022oji,Aloni:2022ebm,MammenAbraham:2023psg,Cheung:2023gwm,Asai:2023xxl} for other studies regarding far-forward neutrinos and new physics.

In the SM, the CC neutrino scatterings off nuclei are driven by the $W$ boson exchange. BSM contributions that could modify these interaction rates are typically associated with new physics at the scale above the characteristic momentum transfer in neutrino interactions at the LHC, especially if they go beyond the SM-like V-A interactions that could be affected by pure neutrino-philic species, cf. Ref.~\cite{Kelly:2021mcd} for sample such analysis for forward LHC searches. Therefore, a convenient way to describe such BSM-induced interactions is via an effective field theory (EFT) approach. The typical momentum transfer in CC DIS neutrino scatterings at the LHC, $Q\sim \mathcal{O}(10~\textrm{GeV})$, and we require the new physics scale to remain higher, $\Lambda\gg Q$, for the validity of the EFT.

The sensitivity reach of FASER$\nu$ to a number of such operators that could arise, e.g., within the framework of the weak EFT~\cite{Jenkins:2017jig,Falkowski:2019xoe,Falkowski:2019kfn}, has been studied in Ref.~\cite{Falkowski:2021bkq} and competitive exclusion bounds have been found for some of them, primarily related to $\nu_\tau$-like CC scattering signatures. Here, for illustration, we focus on two such right-handed operators that are described by the following Lagrangian
\begin{eqnarray}
\mathcal{L} & = & -\frac{2\,V_{ud}}{v^2}\times
 (\bar{u}\gamma^\kappa P_Rd) \times \\ & & \left[\epsilon_R^{\mu\tau}\,(\bar{\ell}_\mu\gamma_\kappa P_L \nu_\tau) + \epsilon_R^{\tau e}\,(\bar{\ell}_\tau\gamma_\kappa P_L \nu_e)\right],\nonumber
\end{eqnarray}
where we use $V_{ud}$ as the relevant entry of the Cabibbo-Kobayashi-Maskawa (CKM) matrix, $v \simeq 246~\textrm{GeV}$ is the SM Higgs vacuum expectation value, and $\epsilon_R^{\alpha\beta}$ are the respective Wilson coefficients describing neutrino NSI. 

The presence of neutrino NSI would affect both production and interaction rates of neutrinos. We follow the discussion of Ref.~\cite{Falkowski:2021bkq} and apply the neutrino detection and production coefficients modified by new physics contributions derived therein. In particular, it has been found that these coefficients are not expected to vary significantly with the incident neutrino energy in the range relevant to the far-forward LHC searches. Hence, they are not strongly sensitive to precise modeling of the neutrino energy spectrum. Still, new physics can lead to distinct features in the LHC data by modifying the spectra for only selected neutrino flavors and parent mesons.

We extend the previous analysis by including the modeling of MC prediction uncertainties, as discussed in \cref{sec:methodology}. The bounds presented below are obtained after profiling over all the nuisance parameters describing the neutrino spectra variations. These variations could \textsl{a priori} surpass the impact of neutrino NSI and should be considered in estimating new physics reach. As we present below, however, this effect does not significantly limit the sensitivity of the FPF experiments, at least for EFT operators selected in our analysis. In our analysis, we consider both the energy and spatial distribution of events in the detectors. For the latter distribution, we consider three radial bins for both FASER$\nu$ and FASER$\nu$2. We focus on the emulsion detectors with the best capabilities to study $\nu_\tau$ interactions.

\begin{figure*}[t]
\includegraphics[width=0.5\textwidth]{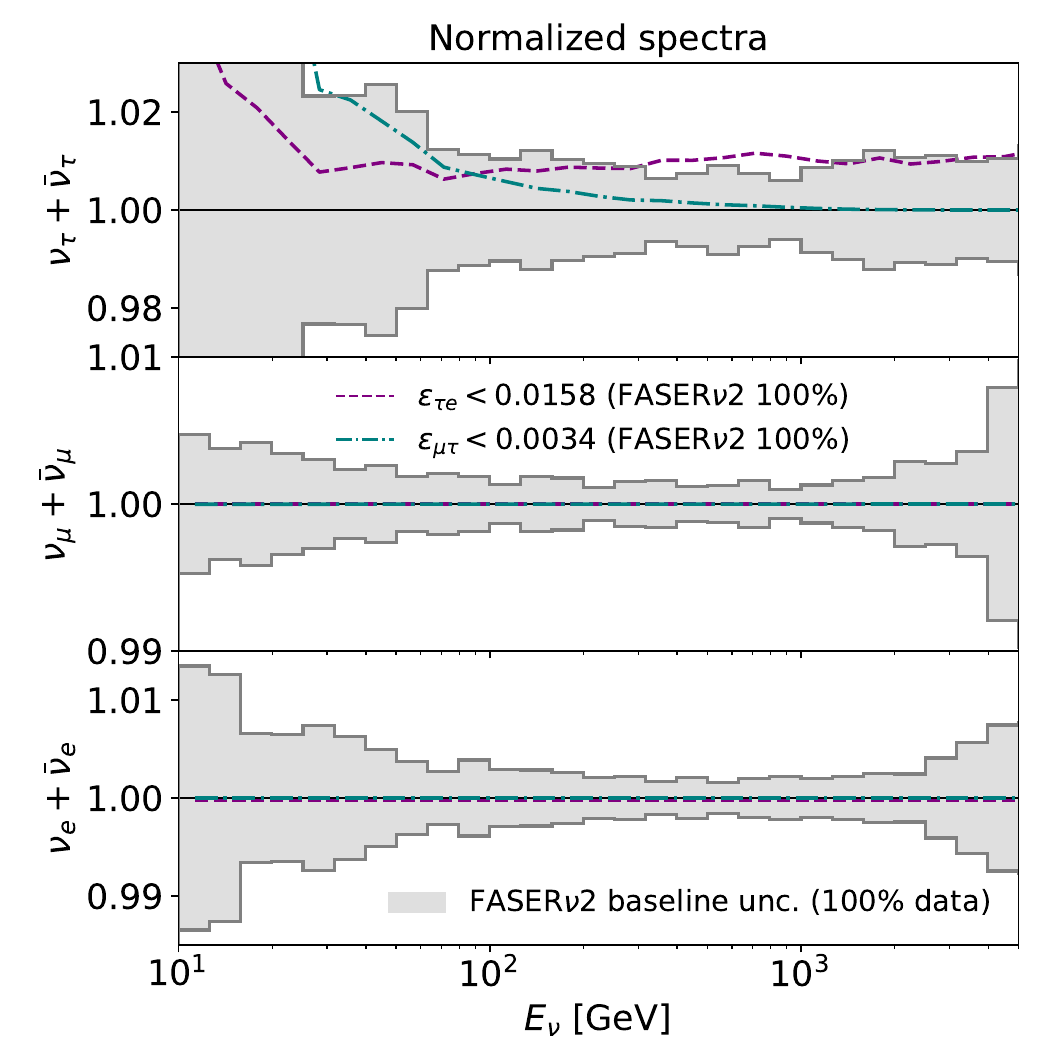}
\includegraphics[width=0.49\textwidth]{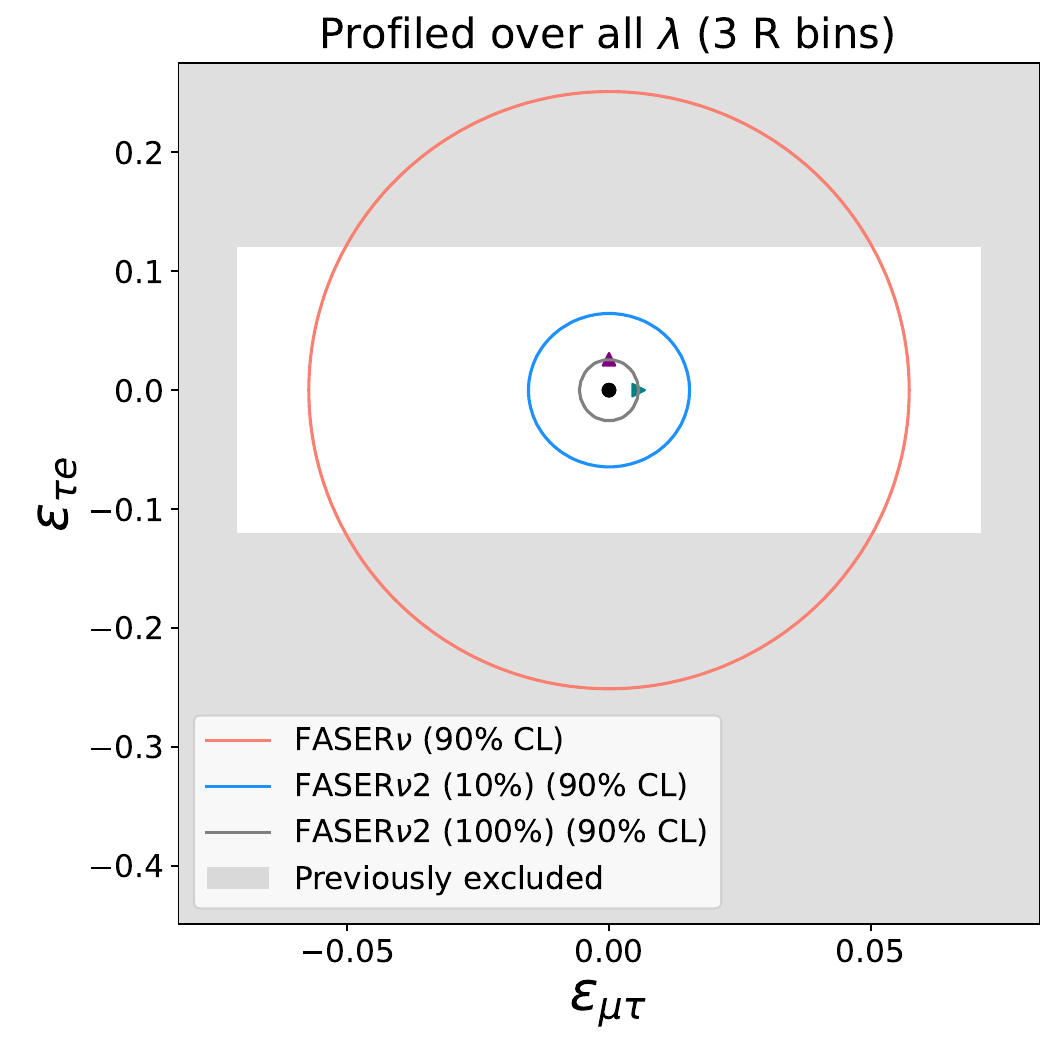}
\caption{\textit{Left:} The uncertainties for the neutrino CC event scattering rates at FASER$\nu$2, assuming 100\% of the data collected and using three radial bins, along with the NSI parameters $\epsilon_{\tau e}$ and $\epsilon_{\mu\tau}$ set to the obtained constraints. \textit{Right:} The projected FASER$\nu$2 constraints are compared to those obtainable using only $10\%$ of the expected data and those attainable with $100\%$ of the expected FASER$\nu$ data. Current bounds on the respective Wilson coefficients are shown with gray-shaded bands.}
\label{mutau_taue_constraints}
\end{figure*}

We present the results of our analysis in \cref{mutau_taue_constraints}. In the left panel, we show gray-shaded uncertainty bands on the electron, muon, and tau-neutrino CC scattering rates in FASER$\nu$2. In this case, no impact of new physics has been assumed. The baseline model is chosen to be an average of the predictions, similar to the results discussed in \cref{sec:spectraconstraints}. On top of this, we also present colorful lines representing predicted deviations from the baseline scenario due to the presence of neutrino NSI. These have been obtained by simultaneously changing both the Wilson coefficients mentioned above and the nuisance parameters describing MC variations. We subsequently profile over all the parameters besides either $\epsilon_R^{\mu\tau}$ or $\epsilon_R^{\tau e}$. 

The former Wilson coefficient $\epsilon_R^{\mu\tau}$ is related to the operator which couples the (charged) muon and tau neutrino. It primarily affects the neutrino production rate by inducing a non-zero branching fraction for the process, $\pi\to\mu\nu_\tau$, which enhances the tau neutrino flux. This operator could also induce CC scatterings of the tau neutrinos leading to the final state muons, $\nu_\tau N\to \mu X$. Such a process would reduce the number of events reconstructed as $\nu_\tau$-like CC scattering interactions, given the lack of the final-state tau lepton $\tau$. However, the net impact on the $\nu_\tau$ production rate, i.e., the increase of the $\nu_\tau$ flux, is significantly more substantial. It is driven by a large flux of parent pions that, otherwise, never produce tau neutrinos.

On the other hand, the Wilson coefficient $\epsilon_R^{\tau e}$ couples $\nu_e$ and the tau lepton $\tau$. In this case, the impact on the $\nu_\tau$-like detection rate is more significant, and it is determined by the NSI-induced CC electron neutrino scatterings, $\nu_e N\to \tau X$, which mimic interactions of the tau neutrinos. The presence of this operator does not induce any additional significant production modes for the electron neutrinos together with the tau lepton. The dominant such modes would be associated with decays of charm hadrons and then related to operators involving quarks from the second generation.  

The projected bounds on both the coefficients considered individually that we obtain at $1\sigma$ and for FASER$\nu$2 read: $|\epsilon_R^{\tau e}|< 0.0158$ and $|\epsilon_R^{\mu\tau}|< 0.0034$. The resulting deviations from the baseline tau neutrino spectrum are at the $\mathcal{O}(1\%)$ level for the $\tau e$ operator, as shown with the purple line in the top left panel of \cref{mutau_taue_constraints}. They do not depend significantly on the incident neutrino energy. This is because the corresponding impact of new physics on the tau neutrino detection rate only mildly depends on $E_\nu$. Instead, in the $\mu \tau$ case, the deviations from the baseline spectrum show clear energy dependence. Notably, in the SM, pion decay contribution to the muon neutrino far-forward spectrum at the LHC dominates at energies below a few hundred GeV. It is then this energy regime in which one expects the most significant enhanced production of $\nu_\tau$s from rare NSI-induced pion decays, which is the reason behind the observed enhanced effect. 

We note that the observation of new physics in interactions at $E_{\nu_\tau} \sim $ few tens of GeV could be affected by a decreasing vertex detection efficiency in emulsion for lowering energies~\cite{FASER:2019dxq}. In order to estimate the impact of this effect on our NSI results, we have additionally studied FASER$\nu$2 bounds after applying the relevant effect and lepton detection efficiency. To this end, we have employed the same efficiency functions as in FASER$\nu$, cf. \cref{sec:detection} for discussion. The projected bounds found this way are about $20\%$ less strong for the $\tau e$ operator. The weakening of the predicted constraints is more pronounced for the $\mu\tau$ operator. The excluded value grows by about $30\%$, as expected from a stronger energy dependence of the NSI effect in this case. In general, however, we find that both operators can be constrained well in FASER$\nu$2 even for decreasing detection efficiency at lower energies. The precise constraining power will be further sensitive to PDF uncertainties, as discussed in \cref{sec:detection}.

In the central and bottom left panels of \cref{mutau_taue_constraints}, we also show with colorful lines the expected NSI-driven deviations from the baseline CC scattering rates for the electron and muon neutrinos. As can be seen, these are significantly smaller than for the tau neutrinos. The observed difference is due to much larger expected scattering rates for $\nu_e$ and $\nu_\mu$ that are less sensitive to small variations in the number of events than $\nu_\tau$s. We note that the results of such analysis would be much different in the presence of non-negligible neutrino oscillations in long-baseline neutrino experiments. Instead, far-forward neutrino searches at the LHC combine capabilities of short-baseline neutrino experiments with the potential to detect $\nu_\tau$-induced CC scattering events directly.

The right panel of \cref{mutau_taue_constraints} corresponds to the results obtained after profiling over all the nuisance parameters but without profiling over both the Wilson coefficients. The projected bounds found this way are similar in constraining power to the ones discussed above. At $90\%$ CL they read $|\epsilon^{\tau e}_R| < 0.026$ and $|\epsilon^{\mu\tau}_R| < 0.0057$. Both considered EFT operators affect the tau neutrino CC event scattering rate almost independently. We also confirm this by finding that the relevant information matrix is close to the diagonal. The expected constraining power of the far-forward neutrino physics program at the LHC can be compared with other searches. In the case of the $\tau e$ operator, the dominant such bounds of $|\epsilon_R^{\tau e}|<0.12$ at $90\%$ CL have been derived in Ref.~\cite{Biggio:2009nt} based on past NOMAD constraints on $\nu_e$ oscillations into $\nu_\tau$~\cite{NOMAD:2001xxt,NOMAD:2003mqg}. The $\mu\tau$ operator can be currently best constrained by using the ratio of pion decay widths to the electron and muon, $\Gamma(\pi\to e\nu_e)/\Gamma(\pi\to\mu\nu_\mu)$~\cite{Bhattacharya:2011qm,ParticleDataGroup:2022pth}. The bounds derived this way are at the level of $|\epsilon_R^{\mu\tau}|<0.071$ at $90\%$ CL~\cite{Falkowski:2021bkq}. As can be seen in the right panel of \cref{mutau_taue_constraints}, the projected FPF bounds can improve past limits by up to an order of magnitude and find new leading limits already with the first $10\%$ of data. We additionally note that in the presence of multiple Wilson coefficients describing non-vanishing neutrino NSI, interesting cancellations can appear that might significantly weaken these bounds in fine-tuned scenarios~\cite{Bhattacharya:2011qm}. In order to better resolve such issues, measuring the final-state neutrino flavor remains crucial, which further highlights the importance of neutrino NSI searches in the FPF experiments.

We also comment on the importance of using double differential distributions in these analyses. Given a relatively small transverse size of both FASER$\nu$ and FASER$\nu$2, we find only mild improvement in using three radial bins over not considering the spatial distribution of events. However, going to larger pseudorapidity regimes could visibly strengthen the bounds. We have numerically studied this by extending the search to $1~\textrm{m}$ away from the beam-collision axis, i.e., to the distance characteristic for FLArE. The proposed AdvSND detector could extend this coverage even further. Based on our analysis, we expect further $\mathcal{O}(10\%)$ improvement in the NSI bounds on $\epsilon_R^{\mu\tau}$ and $\epsilon_R^{\tau e}$ from analyzing the data in the full pseudorapidity range of the FPF experiments.

Finally, it is instructive to comment on an approximate scale of heavy new physics species $\Lambda$, which could be involved in generating the low-energy operators of our interest. This could be obtained by matching our operators to the SMEFT operators above the electroweak (EW) scale~\cite{Cirigliano:2012ab,Jenkins:2017jig,Dekens:2019ept}. In this case, off-diagonal right-handed EFT operators receive only $\Lambda^{-4}$ corrections~\cite{Falkowski:2019xoe}. The FASER$\nu$2 bounds found above could then be translated into about $\Lambda = v/\epsilon^{1/4} \simeq 600~\textrm{GeV}$ and $900~\textrm{GeV}$ at $90\%$CL for the $\tau e$ and $\mu\tau$ operators, respectively. 

\section{Conclusions}
\label{sec:conclusions}

When estimating the discovery potential of a novel experimental program, it always remains crucial to properly consider possible Standard Model effects and related uncertainties that could mimic new phenomena. Breaking this degeneracy is also essential for understanding the expected impact of the recently started far-forward neutrino physics program at the LHC. In the current work, we have made an important step in this direction. 

We have proposed parameterizing the expected neutrino spectra by combining the leading predictions based on various approaches to modeling forward parent hadron spectra. The parameterized flux model obtained this way is characterized by 12 nuisance parameters describing the variations in neutrino spectrum normalization and shape. Importantly, these variations take into account expected correlations between the neutrino spectra of different flavors. We then estimated how well the current and proposed forward LHC neutrino experiments can constrain this model. Our analysis considers information about the neutrino charged-current interaction rates for different flavors, energies, and pseudorapidities. 

In particular, we have shown that the future Forward Physica Facility data will allow for constraining the LHC neutrino fluxes up to even a sub-percent level for $\nu_e$ and $\nu_\mu$, i.e., to precision at which additional PDF uncertainties affecting neutrino interaction rates become important. These will be reduced thanks to future EIC and FPF measurements. The FPF data will then allow for differentiating between various MC predictions with high precision. Instead, the expected uncertainty bands are of order few percent for the tau neutrinos. 

The forward LHC neutrino data will also allow for further improving the tunes of the MC tools used to predict the parent hadron spectra. This will profoundly affect our understanding of cosmic-ray physics, including the possibility of solving the puzzling excess of the muon rate observed in CR-induced air showers at ultrahigh energies. We have analyzed a recently proposed solution to this problem based on the pion-to-kaon swapping among products of high-energy $pp$ collisions at large pseudorapidities. Our study shows that the currently operating FASER$\nu$ detector offers excellent capabilities to probe this scenario within the next few years of LHC Run~3. Future FPF searches could further improve relevant bounds on the swapping fraction up to sub-percent precision.

New physics contributions to neutrino interactions can also be probed this way. We have illustrated this for a $\nu_\tau$-like signature of CC interactions for TeV-scale energies of incident neutrinos. These can be measured on an event-by-event basis in the far-forward emulsion detectors at the LHC. We have tested a scenario in which two Wilson coefficients describing BSM right-handed couplings of quarks to charged leptons and neutrinos are varied simultaneously. We show that the unique effect of new physics can be identified by employing full forward LHC neutrino data to disentangle NSI from variations in MC predictions attributed to an insufficient understanding of the forward hadron production. We have shown that selected Wilson coefficients can be then constrained in the future FASER$\nu$2 detector with up to about an order of magnitude better precision than current bounds.

One can extend the current work to other physics analyses. This includes, i.a., specific effects predicted to modify neutrino production rates, e.g., intrinsic charm~\cite{Brodsky:1980pb,Brodsky:2015fna} or gluon saturation at small $x$~\cite{Gribov:1983ivg,Mueller:1985wy} that will affect the charm-induced tau neutrino spectrum in the far-forward kinematic region. New physics could also non-trivially manifest itself in the LHC neutrino data if oscillations into sterile neutrinos are present~\cite{FASER:2019dxq}, cf. also recent discussion about the discovery prospects for neutrino-modulino oscillations~\cite{Anchordoqui:2023qxv}. The onset of a new era of precision neutrino physics at the LHC offers exciting opportunities to improve our understanding of hadronic interactions and the physics of the most elusive among SM particles.

\section*{Acknowledgements}

We thank Weidong Bai, Atri Bhattacharya, Luca Buonocore, Yu Seon Jeong, Rafal Maciu\l{}a, Mary Hall Reno, Luca Rottoli, Ina Sarcevic, Anna M. Stasto, and Antoni Szczurek for helpful discussions and for sharing the files used to obtain the charm-induced neutrino spectra. We would like to thank Luis Anchordoqui, Akitaka Ariga, Tomoko Ariga, Anatoli Fedynitch, Max Fieg, Tanguy Pierog, Felix Riehn, Dennis Soldin for useful discussions and comments on the manuscript. We are grateful to the authors and maintainers of many open-source software packages, including \textsc{Rivet}~\cite{Buckley:2010ar, Bierlich:2019rhm} and \textsc{scikit-hep}~\cite{Rodrigues:2020syo}. FK acknowledges support by the Deutsche Forschungsgemeinschaft under Germany's Excellence Strategy - EXC 2121 Quantum Universe - 390833306. TM and ST are supported by the National Science Centre, Poland, research grant No. 2021/42/E/ST2/00031. ST is also supported by the grant ``AstroCeNT: Particle Astrophysics Science and Technology Centre" carried out within the International Research Agendas programme of the Foundation for Polish Science financed by the European Union under the European Regional Development Fund. ST is additionally partly supported by the European Union’s Horizon 2020 research and innovation program under grant agreement No 952480 (DarkWave).

\onecolumngrid
\appendix
\onecolumngrid
\section{Application of the Cram$\acute{\textrm{e}}$r-Rao bound to forward LHC neutrino measurements}
\label{sec:informationgeometry}

As discussed in \cref{sec:parameterization}, we interpolate between established predictions for the forward neutrino spectra to obtain the expected number of neutrino interaction events in each of the detectors considered in our study. Here, we discuss further steps of our statistical analysis.

The observables in the binned histogram analysis are the numbers of events $n_i$ observed in each $i$th bin. The likelihood function is obtained as a product of the Poisson likelihoods for all bins
\begin{equation}
L(\text{data}|\text{model})
= \prod_{\text{bins}~i} \text{Pois}(n_i|N_i)
= \prod_{\text{bins}~i} \frac{N_i^{n_i} e^{-N_i}}{n_i!},
\label{PoissonLikelihoods}
\end{equation}
where $N_i$ is the expected number of events per bin in the model. 
In the following, we provide a function for the expected log-likelihood ratio $\log r$, where the likelihood ratio with respect to the baseline model reads
\begin{equation}
   r(\lambda^\pi,\lambda^K,\lambda^c) = \frac{ L(\text{expected data}|\lambda^\pi,\lambda^K,\lambda^c)}{ L(\text{expected data}|\lambda^\pi=0,\lambda^K=0,\lambda^c=0)}
\end{equation}
with the expected data corresponding to $\lambda^\pi=\lambda^K=\lambda^c=0$. 

The expected likelihood ratio is approximated as
\begin{equation}
  -2\log r = - \frac{d^2 \log r}{d\lambda^{(i)} d\lambda^{(j)}} \Delta \lambda^{(i)} \Delta \lambda^{(j)}
           = I_{ij} \Delta \lambda^{(i)} \Delta \lambda^{(j)},
\end{equation}
where $(i),(j)$ run over all parent hadrons $\pi, K, c$ for all generators, and $I_{ij}$ are the components of the Fisher Information matrix. By the Cram{\'e}r--Rao bound~\cite{cramer-rao, cramer-rao2}, the smallest uncertainty achievable in the measurement is then obtained when the covariance matrix $\mathrm{C}_{ij} = I_{ij}^{-1}$. To avoid introducing additional numerical uncertainty in the computation of the Fisher information, the expected number of events per bin in the model is generalized into a real positive parameter in Eq.~\eqref{PoissonLikelihoods}.
The uncertainty bands for the neutrino spectra are obtained by solving for the eigenvalues and -vectors of the information matrix. The model is then varied from the baseline to the direction of each eigenvector individually, and the uncertainty in each bin is obtained as the square root of the quadratic sum of the differences of each variation to the baseline. 
When using multiple radial bins,  
the uncertainty $\delta_i$ for each $i$-th radial bin is first computed in the aforementioned way. These are then combined as 
$\delta_\textrm{tot} = \sqrt{\sum_i \delta_i^2} \left(\sum_i \delta_i \right)^{-1}$, 
separately for all energy bins, yielding the total uncertainty shown in the spectrum plots. 
In the present work, the uncertainties of all spectra are reported at the $1\sigma$ level. Results corresponding to different statistical significance are also provided in selected cases in \cref{sec:applications}.

We use a profiling procedure amounting to a parallel projection of a generalized ellipsoid in the parameter space to estimate the constraints that can be obtained for a parameter used in the model computation. To profile over the $n$-th parameter in the information matrix $I$, the $n$-th column (or row) of $I$, with the $n$-th entry removed, is taken as the vector $\mathbf{m}$ describing the mixing between the profiled parameter and the remainder.
A reduced information matrix $I^{\rm reduced}$ is attained by removing the $n$-th column and row from $I$, and the profiled information matrix is given by~\cite{Brehmer:2016nyr}
\begin{equation}
I^{\rm profiled}
=
I^{\rm reduced}
-
\mathbf{m} \otimes \mathbf{m} / I_{nn}.
\end{equation}
The procedure is repeated to profile over multiple parameters, starting with the information matrix resulting from the previous step. 
By profiling over all but one parameter, the information matrix eventually reduces into a single entry $a$, and the ultimate constraint for the remaining parameter is then obtained as $a^{-1/2}$.

\clearpage
\bibliography{references}

\end{document}